# LOCAL ANTITHETIC SAMPLING WITH SCRAMBLED NETS

By Art B. Owen

*Stanford University*

We consider the problem of computing an approximation to the integral $I = \int_{[0,1]^d} f(x)\,dx$. Monte Carlo (MC) sampling typically attains a root mean squared error (RMSE) of $O(n^{-1/2})$ from $n$ independent random function evaluations. By contrast, quasi-Monte Carlo (QMC) sampling using carefully equispaced evaluation points can attain the rate $O(n^{-1+\varepsilon})$ for any $\varepsilon > 0$ and randomized QMC (RQMC) can attain the RMSE $O(n^{-3/2+\varepsilon})$, both under mild conditions on $f$.

Classical variance reduction methods for MC can be adapted to QMC. Published results combining QMC with importance sampling and with control variates have found worthwhile improvements, but no change in the error rate. This paper extends the classical variance reduction method of antithetic sampling and combines it with RQMC. One such method is shown to bring a modest improvement in the RMSE rate, attaining $O(n^{-3/2-1/d+\varepsilon})$ for any $\varepsilon > 0$, for smooth enough $f$.

**1. Introduction.** Many problems in science and engineering require multidimensional quadratures. There we seek the value of an integral $I = \int_{[0,1]^d} f(x)\,dx$. The integrand $f$ subsumes any transformations necessary to account for noncubic domains, or integration with respect to a nonuniform density. Monte Carlo sampling is often employed for these problems. Its basic form uses an estimate $\hat{I} = (1/n)\sum_{i=1}^n f(x_i)$, where $x_i$ are simulated independent draws from $U[0,1]^d$. When $f$ is in $L^2$, then Monte Carlo has a root mean squared error (RMSE) at the familiar $O(n^{-1/2})$ rate.

Monte Carlo integration can be improved by the use of variance reduction methods. Well-known techniques include stratification, importance sampling, control variates and antithetic sampling. These are described in texts such as Glasserman [10] and Fishman [8].

Received August 2007; revised August 2007.
[1]Supported in part by NSF Grants DMS-03-06612 and DMS-06-04939.

In stratification, the sample points $x_1, \ldots, x_n$ are made more uniformly distributed than they would be by chance. This idea of choosing points more uniformly than they would be by chance underlies quasi-Monte Carlo (QMC) sampling which can be thought of as an extreme version of stratification. Deterministic QMC methods can attain an error rate of $O(n^{-1+\varepsilon})$, while randomized versions can achieve an RMSE of $O(n^{-3/2+\varepsilon})$, both under mild smoothness conditions on $f$, for any $\varepsilon > 0$.

It is interesting to investigate whether variance reduction techniques from MC bring any advantages to the QMC setting. Chelson [3] and Spanier and Maize [27] have investigated QMC with importance sampling. Hickernell, Lemieux and Owen [12] have studied the combination of QMC with control variates. This paper considers a combination of QMC with antithetic sampling.

Antithetic sampling improves Monte Carlo by exploiting spatial structure in $f$. Each point $x \in [0,1]^d$ is coupled with another $\widetilde{x}$, commonly obtained as $\widetilde{x} = 1 - x$ interpreted componentwise. In practice, we average $\widetilde{f}(x_i) = (f(x_i) + f(\widetilde{x}_i))/2$ at $n/2$ points $x_i$. If $f(x)$ is linear in $x$, then $\widetilde{f}(x_i) = I$ and $I$ can be estimated without error. When $f(x)$ is nearly linear or nearly antisymmetric [i.e., $f(x) - I \doteq I - f(\widetilde{x})$], then antithetic sampling can bring a great reduction in RMSE, although the rate remains $n^{-1/2}$. In local antithetic sampling, described below, the point $\widetilde{x}$ is always close to $x$. Since smooth functions are locally linear in the Taylor approximation sense, local antithetic sampling can be much better than antithetic sampling for small $d$.

This paper considers several ways of combining antithetic sampling and randomized digital nets. The main result is that one such method, a box folding scheme, reduces the RMSE to $O(n^{-3/2-1/d+\varepsilon})$. The improvement in rate is modest and diminishes with $d$. But it compares favorably with ordinary antithetic sampling which only changes the constant in the RMSE, and changes it for the worse for some $f$. The other variance reduction methods from MC (control variates and importance sampling) only act on the constant and do not improve the RMSE rate when applied to randomized QMC.

The improvement we find is the same factor $n^{-1/d}$ from classic results of Haber [11]. Haber gets an RMSE rate of $O(n^{-1/2-1/d})$ for cubically stratified sampling and it improves to $O(n^{-1/2-2/d})$ for a locally antithetic version of that sampling.

The outline of this paper is as follows. Section 2 summarizes background information on scrambled nets, which are a form of randomized quasi-Monte Carlo sampling. Section 3 introduces some new notions of $d$-dimensional folding operations used to introduce local antithetic properties into digital nets, and proposes three specific methods. Section 4 illustrates several reflection net sampling schemes on a two-dimensional integrand studied by [25]. The root mean squared errors seem to follow a $n^{-2}$ rate. The next sections are devoted to showing that one of the methods, box folding, attains an RMSE of $O(n^{-3/2-1/d+\varepsilon})$. Section 5 recaps the variance for scrambled net quadrature of smooth functions. It corrects an error in the proof of the $O(n^{-3/2}(\log n)^{(d-1)/2})$ RMSE rate from [21]. It also extends the proof there to a wider collection of digital nets and uses a weaker smoothness condition than the earlier paper had. Section 6 builds on Section 5 to prove that the RMSE of the box folding scheme is $O(n^{-3/2-1/d}(\log n)^{(d-1)/2})$ in $d$ dimensions. More smoothness is required for this result than for the unreflected scrambled nets. Section 7 presents the box folding scheme as a hybrid of a monomial cubature rule with scrambled net sampling. Finally, it discusses how one might make use of these findings in higher dimensional problems of low effective dimension.

**2. Background and notation.** Scrambled nets are a particular form of randomized quasi-Monte Carlo sampling. The monograph [17] by Niederreiter is the definitive source for quasi-Monte Carlo sampling. Randomized quasi-Monte Carlo sampling was surveyed by Lemieux and L'Ecuyer [15]. Scrambled nets were first proposed in [19].

We use superscripts for components, so $x, x_i \in [0,1]^d$ have components $x^j$ and $x_i^j$ respectively for $j = 1, \ldots, d$. The set $\{1, \ldots, d\}$ is abbreviated $1\!:\!d$. If $u \subseteq 1\!:\!d$, then its complement $\{1 \le j \le d \mid j \notin u\}$ is written as $-u$.

We often have to extract and combine components from one or more points in $[0,1]^d$. When we extract the components $x^j$ for $j \in u \subseteq 1\!:\!d$, we use $x^u$ to denote the result. When $x, z \in [0,1]^d$ and we want to combine $x^u$ with $z^{-u}$, we write it as $x^u\!:\!z^{-u}$. Thus, $x^u\!:\!z^{-u}$ is the point $y \in [0,1]^d$ with $y^j = x^j$ for $j \in u$ and $y^j = z^j$ for $j \notin u$.

2.1. *Quasi-Monte Carlo.* Like plain Monte Carlo, quasi-Monte Carlo sampling estimates an integral $I = \int_{[0,1]^d} f(x)\,dx$ by the average $\hat{I} = \frac{1}{n}\sum_{i=1}^n f(x_i)$ taken over points $x_i \in [0,1]^d$. QMC aims to be better than random by selecting $x_i$ to be even more uniformly distributed than random points typically are. To quantify the nonuniformity of $x_1, \ldots, x_n$, consider the local discrepancy function

$$\delta(x) = \frac{1}{n}\sum_{i=1}^n 1_{x_i \in [0,x]} - \text{Vol}([0,x]) \tag{1}$$

for $x \in [0,1]^d$. The star discrepancy of $x_1, \ldots, x_n$ is

$$D_n^*(x_1, \ldots, x_n) = \sup_{x \in [0,1]^d} |\delta(x)|. \tag{2}$$

When $d = 1$, then $D_n^*$ reduces to the Kolmogorov–Smirnov distance between the empirical distribution of $x_i$ and the $U[0,1]$ distribution. The Koksma–Hlawka inequality [13] is

$$(3) \qquad |\hat{I} - I| \leq D_n^*(x_1, \ldots, x_n) \|f\|_{\mathrm{HK}},$$

where $\|f\|_{\mathrm{HK}}$ is the total variation of $f$ in the sense of Hardy and Krause. It is possible to construct $x_i$ so that $D_n^* \leq C_d (\log n)^{d-1}/n$ for $n > 1$. With such constructions, $|\hat{I} - I| = O(n^{-1+\varepsilon})$ holds for all $\varepsilon > 0$, under the mild condition that $\|f\|_{\mathrm{HK}} < \infty$. Thus, QMC has a far better asymptote than MC.

2.2. *Digital nets.* Digital nets attain their low discrepancy by being simultaneously stratified for many different stratifications of $[0,1]^d$. Those stratifications are defined through hyper-rectangular subsets known as elementary intervals.

This section defines these elementary intervals and some digital nets and digital sequences. Throughout we use $b$ to denote an integer base in which to represent real numbers, $d$ to represent the dimension, $k_j$ to represent some nonnegative integer powers of $b$ and $t_j$ to represent some nonnegative integer translations.

DEFINITION 1. Let $b \geq 2$ and $d \geq 1$ be integers. Let $\kappa = (k_1, \ldots, k_d)$ and $\tau = (t_1, \ldots, t_d)$ be $d$-vectors of integers for which $k_j \geq 0$ and $0 \leq t_j < b^{k_j}$. Then the set

$$\mathbb{B}_{\kappa, \tau} = \prod_{j=1}^d \left[ \frac{t_j}{b^{k_j}}, \frac{t_j + 1}{b^{k_j}} \right)$$

is a base $b$ elementary interval.

If one fixes $\kappa$ and varies $\tau$, the sets $\mathbb{B}_{\kappa, \tau}$ provide a tiling of $[0,1)^d$. The tilings of the three illustrations in Figure 1 are of this type.

The volume of $\mathbb{B}_{\kappa, \tau}$ is $b^{-|\kappa|}$, where $|\kappa| = k_1 + \cdots + k_d$. The closure of $\mathbb{B}_{\kappa, \tau}$, defined by replacing the half open intervals in Definition 1 by closed intervals, is denoted $\overline{\mathbb{B}}_{\kappa, \tau}$. The center of $\mathbb{B}_{\kappa, \tau}$ and of $\overline{\mathbb{B}}_{\kappa, \tau}$ is the point $\mathsf{c}_{\kappa, \tau}$ with $\mathsf{c}^j_{\kappa, \tau} = (t_j + 1/2)/b^{k_j}$.

When one or more of the $k_j$ is 0, then the corresponding factors of $\mathbb{B}$ reduce to $[0, 1)$. Let $u \subseteq 1\!:\!d$ and let $\kappa$ be a vector of length $|u|$ indexed by $j \in u$, with component $k_j$ for $j \in u$. Similarly, let $\tau$ have components $t_j$ for $j \in u$. Then

$$\mathbb{B}_{u, \kappa, \tau} \equiv \prod_{j \in u} \left[ \frac{t_j}{b^{k_j}}, \frac{t_j + 1}{b^{k_j}} \right) \prod_{j \notin u} [0, 1)$$

will be used below. The center of $\mathbb{B}_{u,\kappa,\tau}$ is the point $\mathsf{c}_{u,\kappa,\tau}$ with

$$\mathsf{c}^j_{u,\kappa,\tau} = \begin{cases} \dfrac{t_j + 1/2}{b^{k_j}}, & j \in u, \\ \dfrac{1}{2}, & j \notin u. \end{cases}$$

The elementary interval $\mathbb{B}_{\kappa,\tau}$ in Definition 1 has volume $b^{-|\kappa|}$. Ideally it should get $nb^{-|\kappa|}$ of the sample points $x_1, \ldots, x_n$. If that happens for one vector $\kappa$, we have a stratified sample with one stratum for each $\tau$. Digital nets attain such stratification for multiple $\kappa$ simultaneously.

DEFINITION 2. For integers $m \geq q \geq 0$, $b \geq 2$ and $d \geq 1$, a sequence of points $x_1, \ldots, x_{b^m} \in [0,1)^d$ is a $(q, m, d)$-net in base $b$ if every base $b$ elementary interval in $[0,1)^d$ of volume $b^{q-m}$ contains precisely $b^q$ points of the sequence.

The parameter $q$ defines the quality of the net, with smaller values implying better equidistribution, and $q = 0$ being the very best when it is attainable. The minT system [24] identifies the best known nets (smallest $q$) given the values of $m$, $d$ and $b$. The net property is enough to ensure low discrepancy:

THEOREM 1. *If $x_1, \ldots, x_n$ are a $(q, m, d)$-net in base $b$, then*

$$n \times D_n^*(x_1, \ldots, x_n) \leq \frac{1}{(d-1)!}\left(\frac{\lfloor b/2 \rfloor}{\log b}\right)^{d-1}(\log n)^{d-1} + O(b^q (\log n)^{d-2})$$

*for $n > 1$, where the implied constant in the error term depends only on $b$ and $d$.*

PROOF. This is from Theorem 4.10 of [17]. The multiple of $(\log n)^{d-1}$ can be reduced somewhat when $d = 2$ and $b$ is even, or when $d = 3, 4$ and $b = 2$. □

Some constructions of digital nets are extensible. They let us increase $n$, keeping the stratification property and retaining the earlier function evaluations.

DEFINITION 3. For integers $q \geq 0$, $b \geq 2$, and $d \geq 1$, an infinite sequence of points $x_i \in [0,1)^d$ for $i \geq 1$ is a $(q, d)$-sequence in base $b$ if every subsequence $x_{rb^m+1}, \ldots, x_{rb^m+b^m}$, for integers $m \geq q$ and $r \geq 0$, is a $(q, m, d)$-net in base $b$.

It is convenient to work with the first $n = \lambda b^m$ points of the sequence. Should they prove inadequate, one can increase $\lambda$ or, more generally, use $\widetilde{n} = \widetilde{\lambda} b^{\widetilde{m}} \geq n$. The points of the new larger rule include all those of the previous rule. Thus, $(q, d)$-sequences provide extensible integration rules. They automatically satisfy the $(\lambda, q, m, d)$-net property:

DEFINITION 4. For integers $m \geq q \geq 0$, $b \geq 2$, $1 \leq \lambda < b$ and $d \geq 1$, a sequence of points $x_1, \ldots, x_{\lambda b^m} \in [0, 1)^d$ is a $(\lambda, q, m, d)$-net in base $b$ if every base $b$ elementary interval in $[0, 1)^d$ of volume $b^{q-m}$ contains precisely $\lambda b^q$ points of the sequence and no $b$-ary box in $[0, 1)^d$ of volume $b^{q-m-1}$ contains more than $b^q$ points of the sequence.

A *relaxed* $(\lambda, q, m, s)$-net in base $b$ is as above, except that $\lambda \geq b$ is allowed and boxes of volume $b^{q-m-1}$ may have more than $b^q$ points of the sequence.

2.3. *Random digital scrambles.* In scrambled digital net quadrature we take a digital net $a_1, \ldots, a_n \in [0, 1]^d$ and apply a randomizing transformation to this ensemble to produce points $x_1, \ldots, x_n \in [0, 1]^d$ with two properties: each $x_i$ is individually $U[0, 1]^d$ distributed, and $x_1, \ldots, x_n$ are collectively a digital net with probability 1. The first property makes the sample average $\hat{I} = \frac{1}{n} \sum_{i=1}^n f(x_i)$ an unbiased estimate of $I$. The second property means that $\hat{I}$ inherits the good accuracy properties of digital nets.

Some such randomized nets were presented in [19] where it was also shown that scrambled digital sequences remain digital sequences with probability one. The original motivation for randomizing nets was that it allowed independent replications for the purposes of estimating error. That randomization can improve the error rate was at first a surprise, but is now understood as an error cancellation phenomenon.

Randomizations of nets typically use the same random procedure on each point $a_i$ in order to yield the corresponding $x_i$, and so we need only describe the randomization of a single point $a \in [0, 1]^d$. Furthermore, the randomizations applied to components $a^1$ through $a^j$ are typically chosen to be statistically independent. And so we only need to describe the randomization of a single point $a \in [0, 1]$.

It is beyond the scope of this article to explain how randomization of nets is able to achieve the two defining properties. For that one can consult the proposal of Owen [19], it's derandomization by Matoušek [16], and the survey of Lemieux and L'Ecuyer [15]. We can, however, look at the mechanics of some randomizations.

To scramble the point $a \in [0, 1)$, we first write it out in base $b$ as $a = \sum_{k=1}^{\infty} a_{(k)} b^{-k}$, where $a_{(k)} \in \{0, 1, \ldots, b-1\}$. Some values of $a$ have two representations, one ending in infinitely many zeros and the other ending in $b-1$'s.

In such cases we use the representation ending in zeros. For this reason we do not scramble the value $a = 1$, and so scrambled nets actually produce points $x_i \in [0,1)^d$ from points $a_i \in [0,1)^d$. This presents no problem. The standard net constructions yield points in $[0,1)^d$ and $\int_{[0,1)^d} f(x)\,dx = \int_{[0,1]^d} f(x)\,dx$.

The scrambled version of $a$ is the point $x = \sum_{k=1}^{\infty} x_{(k)} b^{-k}$ for digits $x_{(k)} \in \{0, 1, \ldots, b-1\}$ obtained by random permutation schemes applied to the $a_{(k)}$. In practice, the expansion of $x$ is truncated.

There are $b!$ distinct permutations of $\{0, 1, \ldots, b-1\}$. In a uniform random permutation of this set, each permutation has probability $b!$. The method in [19] uses a great many uniform random permutations to scramble $a$. One permutation is applied to the first digit yielding $x_{(1)} = \pi_1(a_{(1)})$. For the $k$th digit $a_{(k)}$, one of $b^{k-1}$ independent uniform random permutations is used to make $x_{(k)}$, chosen based on the value of $\lfloor b^{k-1} a \rfloor$.

The original randomization is computationally burdensome, requiring considerable storage. Matoušek [16] found an alternative and less costly scrambling, by derandomization. We describe that and several other scramblings here. Some more scramblings are described in [23] from which the permutation and scrambling nomenclature used here is taken.

DEFINITION 5. If $b$ is a prime number, then a linear random permutation of $\{0, 1, \ldots, b-1\}$ has the form $\pi(a) = h \times a + g \bmod b$, where $h \in \{1, \ldots, b-1\}$ and $g \in \{0, 1, \ldots, b-1\}$ are independent random variables uniformly distributed over their respective ranges.

Linear permutations are restricted to prime $b$ because otherwise there are nonzero $h$ for which $h \times a + g$ is not a permutation. For example, consider $b = 4$ and $h = 2$. Linear permutations have a generalization, via Galois field arithmetic, to bases that are prime powers, but we do not use them here.

DEFINITION 6. For a prime base $b$, an affine matrix scramble takes the form
$$x_{(k)} = C_k + \sum_{j=1}^{k} M_{kj} a_{(j)} \mod b,$$
where $C_k$ and $M_{kj}$ are in $\{0, 1, \ldots, b-1\}$.

We will consider affine matrix scrambles in which the $C_k$ are independent uniformly distributed elements of $\{0, 1, \ldots, b-1\}$, independent of the elements $M_{kj}$. Such scrambles always have $x \sim U[0,1]$ regardless of $a$ and $M_{kj}$.

The matrix scrambles we consider differ in the structure of the matrix $M$. In each case $M$ is lower triangular and invertible. Invertibility is required so that distinct points $a$ lead to distinct points $x$. The structures that we consider for $M$ can be represented as

(4)
$$\begin{pmatrix} h_1 & & & & \\ g_{21} & h_2 & & & \\ g_{31} & g_{32} & h_3 & & \\ g_{41} & g_{42} & g_{43} & h_4 & \\ \vdots & \vdots & \vdots & \vdots & \ddots \end{pmatrix}, \begin{pmatrix} h_1 & & & & \\ g_2 & h_1 & & & \\ g_3 & g_2 & h_1 & & \\ g_4 & g_3 & g_2 & h_1 & \\ \vdots & \ddots & \ddots & \ddots & \ddots \end{pmatrix},$$
$$\begin{pmatrix} h_1 & & & & \\ h_1 & h_2 & & & \\ h_1 & h_2 & h_3 & & \\ h_1 & h_3 & h_3 & h_4 & \\ \vdots & \vdots & \vdots & \vdots & \ddots \end{pmatrix},$$

where $h$'s are sampled from $\{1, 2, \ldots, b-1\}$ and $g$'s are sampled from $\{0, 1, \ldots, b-1\}$. Within each matrix, entries with the same symbol are identical and entries with different symbols are sampled independently. The matrices in (4) describe respectively, random linear scrambling of [16], $I$-binomial scrambling of [30] and affine striped matrix (ASM) sampling from [23].

Random linear scrambling leads to the same sampling variance as the original net scrambling in [19] (called "nested uniform scrambling") but requires much less storage. $I$-binomial scrambling also leads to the same sampling variance but does so with still less storage.

The ASM scrambling is not variance equivalent to nested uniform scrambling. In the case $d = 1$, ASM attains an RMSE of $O(n^{-2})$, when $f''(x)$ is bounded, which is better than the rate $O(n^{-3/2})$ from other scrambles, though not as good as the rate $O(n^{-5/2})$ that Haber's method gets for $d = 1$.

Our strategy for improving randomized nets is to build in directly some $d$-dimensional versions of locally antithetic sampling. The local antithetic sampling strategy is implemented by adjoining to the scrambled net certain reflections of sample points.

2.4. *ANOVA*. For a function $f \in L^2[0,1]^d$, the ANOVA decomposition is available to quantify the extent to which $f$ depends primarily on lower dimensional projections of the input space. Informally it is like embedding a regular $K^d$ grid in $[0,1]^d$, running an ANOVA on that grid and letting $K \to \infty$. The ANOVA of $[0,1]^d$ was introduced by Hoeffding [14], figures in the Efron–Stein inequality [6], and was independently discovered by Sobol' [26]. For more details and the early history of the ANOVA decomposition, see [29].

We write $f(x) = \sum_{u \subseteq 1:d} f_u(x)$, where $f_u(x)$ is a function of $x$ that depends on $x$ only through $x^u$. To get $f_u$, we subtract strict sub-effects $f_v$ for $v \subsetneq u$ and then average the residual over $x^{-u}$. Specifically,

$$f_u(x) = \int f(x)\,dx^{-u} - \sum_{v \subsetneq u} f_v(x). \qquad (5)$$

The ANOVA terms are orthogonal in that $\int f_u(x) f_v(x)\,dx = 0$ for subsets $u \ne v$. Letting $\sigma_u^2 = \int f_u(x)^2\,dx$, we find that $\sigma^2 = \sum_{|u|>0} \sigma_u^2$.

2.5. *Smoothness and mixed partial derivatives.* This section introduces our notion of smoothness for $f$ and records some elementary consequences of the definition for later use. The mixed partial derivative of $f$ taken once with respect to $x^j$ for each $j \in u$ is denoted by $\partial^u$ with the convention that $\partial^\varnothing f(x) = f(x)$.

DEFINITION 7. *The real valued function $f(x)$ on $[0,1]^d$ is smooth if $\partial^u f(x)$ is continuous on $[0,1]^d$ for all $u \subseteq 1{:}d$.*

REMARK 1. There are $|u|!$ orders in which the mixed partial derivative $\partial^u f(x)$ can be interpreted. The continuity conditions in Definition 7 are strong enough to ensure that all orderings give the same function.

LEMMA 1. *If $f$ is smooth, then $\partial^u f_u(x)$ is continuous for all $u \subseteq 1{:}d$.*

PROOF. The details are omitted to save space. The key is to prove by induction on $|u|$ that $\partial^u \int f(x)\,dx^{-u} = \int \partial^u f(x)\,dx^{-u}$. □

We also need a version of the fundamental theorem of calculus. For points $a, b \in [0,1]^d$, define their rectangular hull as the Cartesian product

$$\mathrm{rect}[a,b] = \prod_{j=1}^d [\min(a_j, b_j), \max(a_j, b_j)].$$

For $d = 1$, if $f$ has a continuous derivative $f'$ on the interval $\mathrm{rect}[c,x]$, then $f(x) = f(c) + \int_{[c,x]} f'(y)\,dy$, with the interpretation that $\int_{[c,x]}$ means $-\int_{[x,c]}$ when $c > x$. For general $d$ and smooth $f$, we have

$$f(x) = \sum_{u \subseteq \{1,\dots,d\}} \int_{[c^u, x^u]} \partial^u f(c^{-u} : y^u)\,dy^u. \qquad (6)$$

Here $\int_{[c^u, x^u]}$ denotes $\pm \int_{\mathrm{rect}[c^u, x^u]}$ where the sign is negative if and only if $c^j > x^j$ holds for an odd number of indices $j \in u$. The term for $u = \varnothing$ equals $f(c)$ under a natural convention.

More generally, let $w \subseteq \{1,\dots,d\}$ and suppose that $\partial^u f$ is continuous for $u \subseteq w$. Then

$$f(x) = \sum_{u \subseteq w} \int_{[c^u, x^u]} \partial^u f(x^{-w} : c^{w-u} : y^u) \, dy^u. \tag{7}$$

For $v \subseteq u \subseteq \{1,\dots,d\}$, let $\partial^{u,v} f$ denote the partial derivative of $f^u$ taken once with respect to each $x^j$ for $j \in v$. That is, $f^{u,v}$ is $f$ differentiated with respect to $x^j$ twice for $j$ in $v$ and once for $j$ in $u - v$.

DEFINITION 8. The real valued function $f(x)$ on $[0,1]^d$ is doubly smooth if $\partial^{u,v} f(x)$ is continuous on $[0,1]^d$ for all $v \subseteq u \subseteq 1:d$.

**3. $b$-ary reflections and folds.** Antithetic sampling is implemented via reflections about the center point of $[0,1]^d$. To induce various local antithetic properties, we will use reflections of a point $x$ about the center of an elementary interval containing $x$.

The case $d = 1$ is simplest. The point $x \in [0,1)$ belongs to the interval $[tb^{-k}, (t+1)b^{-k})$, where $t = t(x) = \lfloor b^k x \rfloor$. The center of this interval is $c = c_k(x) = (t + 1/2)b^{-k}$. The $k$th order reflection of $x$ is $\mathcal{R}_k(x) = 2c_k(x) - x$. The value $k = 0$ corresponds to the simple reflection $1 - x$.

If the base $b$ expansion of $x \in [0,1)$ is $x = \sum_{\ell=1}^{\infty} x_{(\ell)} b^{-\ell}$ with each $x_{(\ell)} \in \{0, 1, \dots, b-1\}$, using trailing 0's when $x$ has two base $b$ representations, then

$$\mathcal{R}_k(x) = \sum_{\ell=1}^{k} x_{(\ell)} b^{-\ell} + \sum_{\ell=k+1}^{\infty} (b - 1 - x_{(\ell)}) b^{-\ell}. \tag{8}$$

The reflection $\mathcal{R}_k$ leaves the first $k$ digits of $x$ unchanged and it flips the trailing digits.

By convention, we take $\mathcal{R}_k(1) = \lim_{x \to 1} \mathcal{R}_k(x) = 1 - 1/b^k$. Under this convention we find that $\lim_{k \to \infty} \mathcal{R}_k(x) = x$ holds uniformly in $x$. The reflection is nearly idempotent because $\mathcal{R}_k(\mathcal{R}_k(x)) = x$ unless $x = tb^{-k}$ for an integer $t$ with $0 \le t < b^k - 1$. Note that a reflection of a reflection is not generally a reflection. For instance, when $x$ is not of the form $tb^{-k}$, then $\mathcal{R}_7(\mathcal{R}_3(x))$ flips digits 4 through 7 inclusive of $x$ and leaves all other digits unchanged.

It is useful to consider transformations in which some components of $x$ are reflected, while others get an identity transformation. For simplicity, we adopt the special value $k = -1$, sometimes displayed simply as $-$, to denote the identity transformation, so that $\mathcal{R}_{-1}(x) = x$ for $x \in [0,1]$.

DEFINITION 9. For the vector $\kappa = (k_1, \dots, k_d)$ with $k_j \in \{-1, 0, 1, \dots\}$, the reflection $\mathcal{R}_\kappa$ of $x \in [0,1]^d$ is defined by

$$\mathcal{R}_\kappa(x) = z \in [0,1]^d, \qquad \text{where } z^j = \mathcal{R}_{k_j}(x^j). \tag{9}$$

Figure 1 illustrates some reflections $\mathcal{R}_{(1,2)}$ and $\mathcal{R}_{(-,2)}$ for $x \in [0,1)^2$ with $b=2$, as well as a box fold described below. Geometrically, a reflection of $x$ has some components symmetric about the center of an elementary interval containing $x$ and all other components equal to the corresponding ones of $x$.

Recall that the center of the elementary interval $\mathbb{B}_{\kappa,\tau}$ is the point

$$\tag{10} \mathsf{c}_{\kappa,\tau} = \left( \frac{t_1 + 1/2}{b^{k_1}}, \ldots, \frac{t_d + 1/2}{b^{k_d}} \right).$$

For a vector $\kappa = (k_1, \ldots, k_d)$ with $k_j \geq 0$, the point $x \in [0,1)^d$ belongs to the elementary interval $\mathbb{B}_{\kappa,\tau}$ for $\tau = \tau(\kappa, x) = \lfloor b^\kappa x \rfloor$, with the multiplication and floor operators taken componentwise. For such $\kappa$, the reflection $\mathcal{R}_\kappa(x)$ may be written

$$\mathcal{R}_\kappa(x) = 2\mathsf{c}_{\kappa,\tau(\kappa,x)} - x.$$

Notice that $\mathcal{R}_\kappa(x)$ has some points of discontinuity whenever $\max_j k_j \geq 1$ because then $\mathsf{c}_{\kappa,\tau(\kappa,x)}$ jumps when $x$ crosses the boundary of certain base $b$ elementary intervals.

DEFINITION 10. Let $x_1, \ldots, x_n \in [0,1)^d$ and let $\mathcal{R}_\kappa$ be a $b$-ary reflection. The folded sequence $\mathcal{F}_\kappa(x_1, \ldots, x_n)$ is the sequence $z_1, \ldots, z_{2n} \in [0,1)^d$ with $z_i = x_i$ for $i = 1, \ldots, n$ and $z_i = \mathcal{R}_\kappa(x_{i-n})$ for $i = n+1, \ldots, 2n$.

If $\mathcal{F}_\kappa(\mathcal{F}_{\kappa'}(x_1, \ldots, x_n))$ and $\mathcal{F}_{\kappa'}(\mathcal{F}_\kappa(x_1, \ldots, x_n))$ are both well defined, then they both have the same points, but possibly in a different order. In this sense, folding is commutative. If $r$ folds have been applied, then the sample size is $2^r n$, perhaps including some points multiple times.

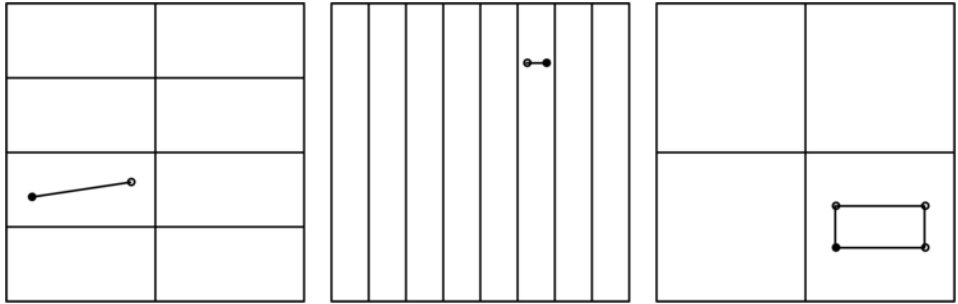

FIG. 1. This figure illustrates some base $b=2$ digital reflections as described in the text. The left panel shows 8 elementary intervals, one of which contains a solid point with its $\mathcal{R}_{(1,2)}$ reflection. The center panel shows 8 elementary intervals, one of which has a point with its $\mathcal{R}_{(3,-)}$ reflection. The right panel shows 4 elementary intervals, one of which includes a solid point $x$ with the other three points of its box reflection $\mathcal{F}_{(1,-)}(\mathcal{F}_{(-,1)}(x))$.

For folding to improve on a digital net, it should produce a local antithetic property within elementary intervals of volume comparable to $b^{q-m}$. To see why, consider the alternatives, taking $q = 0$ for simplicity. If reflections take place within elementary intervals of volume $b^{-r} \ll b^{-m}$, then some elementary intervals of volume $b^{-r}$ have two nearly identical sampling points, while most have none. Conversely, reflections within elementary intervals of volume $b^{-r} \gg b^{-m}$ are not "local enough" to get the best error rate. In particular, if $r$ is constant while $m \to \infty$, then one cannot expect an improved convergence rate, though the leading constant might be better than without folding.

For $\kappa = (k_1, \ldots, k_d)$ with $k_j \in \{-1, 0, 1, \ldots\}$, let $\kappa^+$ have components $k_j^+ = \max\{k_j, 0\}$ and put $|\kappa^+| = \sum_{j=1}^d k_j^+$. Then for $x \in \mathbb{B}_{\kappa^+, \tau}$ of volume $b^{-|\kappa^+|}$, $\mathcal{R}_\kappa(x)$ is in the closed elementary interval $\overline{\mathbb{B}}_{\kappa^+, \tau}$. For reflections of a digital net, we should use $\kappa$ with $|\kappa^+|$ close to $m - q$. When the reflections get finer as $m$ increases, then the reflected scrambled nets will not ordinarily be extensible.

Here we present three methods for inducing local antithetic properties in some $(q, m, 2)$-nets. They are given in increasing order with respect to the number of reflections required.

### 3.1. Reflection nets.

The reflection net takes the form $\mathcal{F}_\kappa(x_1, \ldots, x_n)$, where $x_1, \ldots, x_n$ is a $(\lambda, q, m, d)$-net in base $b$ and $\kappa$ is a vector of $d$ nonnegative integers summing to $q - m$. The reflection net is a (relaxed) $(2\lambda, q, m, d)$-net in base $b$.

For $d = 2$ and $q = 0$, we use $\kappa = (k_1, k_2)$, where each $k_j \doteq m/2$, specifically,

$$\tag{11} k_1 = \left\lfloor \frac{m+1}{2} \right\rfloor \quad \text{and} \quad k_2 = m - k_1.$$

These reflections treat each component of $x$ nearly equally, and reflect within elementary intervals of volume $1/n$.

### 3.2. Box folded nets.

The asymptotic error of scrambled net quadrature from [21] is governed by the norm of the mixed partial derivative $\partial^{1:d} f$. The reflection net may be thought of as averaging the function $\widetilde{f}(x) = (f(x) + f(\mathcal{R}_\kappa(x)))/2$ over a sample of $n$ values of a scrambled net. The function $\widetilde{f}(x)$ has a mixed partial derivative almost everywhere, when $f$ does. If $j \in u$, then $\partial \mathcal{R}_\kappa(x^j)/\partial x^j = -1$ at almost all points, and we find that mixed partial derivatives of $\widetilde{f}$ of odd order largely cancel, while those of even order are averaged. For $d = 2$, the dominant term in the error comes from $\partial^{\{1,2\}} f$, which is of even order and so does not cancel. Therefore, we consider another scheme that averages

$$\widetilde{f}(x) = \tfrac{1}{4}(f(x) + f(\mathcal{R}_{(k_1, -)}(x)) + f(\mathcal{R}_{(-, k_2)}(x)) + f(\mathcal{R}_{(k_1, k_2)}(x))),$$

over $n$ points, with $k_1$ and $k_2$ as in (11). To construct these points, we apply two folds as in $\mathcal{F}_{(k_1,-)}(\mathcal{F}_{(-,k_2)}(x_1,\ldots,x_n))$. The image $\mathcal{F}_{(k_1,-)}(\mathcal{F}_{(-,k_2)})(x)$ is made up of 4 points, symmetric about the center of a box containing $x$. One such quadruple is shown in Figure 1.

### 3.3. Monomial nets.
A greedier reflection strategy folds together all of

$$\mathcal{R}_{(0,m)}, \mathcal{R}_{(1,m-1)}, \qquad \mathcal{R}_{(2,m-2)}, \ldots, \mathcal{R}_{(m,0)}.$$

When these $m+1$ folds are applied to a $(0,m,2)$-net in base $b$, the resulting points correctly integrate any $f$ that is a sum of piece-wise linear functions linear within elementary intervals of volume $b^m$ or larger. Such "monomial nets" extend the local antithetic property of Haber's stratification schemes to all elementary intervals of volume $b^{-m}$, not just those from one vector $\kappa$. The cost is that the sample size is multiplied by $2^{m+1}$, going from $b^m$ to $2(2b)^m$. When $b=2$ the cost is $2n^2$ function evaluations instead of $n$. For $b>2$, the cost grows superlinearly in $n$, but more slowly than the square of $n$:

$$2(2b)^m = 2(2b)^{\log_b(n)} = 2^{1+\log_b(n)} n = 2n^{1+\log_b(2)}.$$

## 4. Example from Sloan and Joe.
To illustrate the three locally antithetic strategies for nets, we consider an integrand studied by Sloan and Joe [25],

$$g(x) = x^2 \exp(x^1 x^2), \qquad x = (x^1, x^2) \in [0,1]^2.$$

This function is bounded and has infinitely many continuous derivatives. We can expect it to have all the smoothness that any of the reflection techniques discussed above might be able to exploit. Also, there are no symmetries or antisymmetries that would make reflection methods exact for this function.

This function has mean $I = \int_0^1 \int_0^1 g(x^1, x^2) \, dx^1 dx^2 = e - 2$, and variance $\sigma^2 = (3-e)(7e-11)/8$. Using Mathematica, one can find that the ANOVA mean squares for the main effects are

$$\sigma^2_{\{1\}} = \tfrac{1}{3}((10-e)e - 15 + 2\text{Ei}(1) - 2\text{Ei}(2) + \log(4))$$

and

$$\sigma^2_{\{2\}} = (3-e)(e-1)/2,$$

where Ei is the exponential integral function, $\text{Ei}(z) = -\int_{-z}^{\infty} t^{-1} e^{-t} \, dt$. The relative variances (sensitivity indices) of the ANOVA terms are

$$\frac{\sigma^2_{\{1\}}}{\sigma^2} \doteq 0.0729, \qquad \frac{\sigma^2_{\{2\}}}{\sigma^2} \doteq 0.8561 \quad \text{and} \quad \frac{\sigma^2_{\{1,2\}}}{\sigma^2} \doteq 0.0710.$$

This function has a meaningfully large bivariate term accounting for about 7.1 percent of the variance, and so it is not a nearly additive function.

For this paper, we consider a scaled version of $g$, namely,

$$f(x) = \frac{x^2 \exp(x^1 x^2)}{e - 2}, \qquad x = (x^1, x^2) \in [0,1]^2. \tag{12}$$

With this scaling, $\int f(x)\,dx = 1$ and so absolute and relative errors coincide.

All of the integration techniques we consider here are based on the construction of $(0,m,2)$-nets given by Faure [7]. The bases used were $b = 2, 3, 5, 7$. The points were either unscrambled, ASM scrambled, or given a random linear scrambling. Nested uniform and I-Binomial scrambling were not tried because they have the same variance as random linear scrambling. For each base and scrambling method, reflection nets, box nets and monomial nets were tried.

The monomial nets did not perform very well, most likely because of the superlinear (in $n$) sample size that they required. In some instances they were slightly better than the original $(0,m,2)$-nets, but not nearly as good as the other methods. For the other methods, over values of $n$ up to the first power of $b$ larger than 2000, the base 2 methods were almost always the best. Accordingly, we work with $b = 2$ and then extend the computations out to $n = 2^{17}$. For methods with reflections, the sample sizes go out to $2^{18}$, while for box folds, the sample sizes go to $2^{19}$.

Figure 2 shows the error for this function with the methods described above. For deterministic methods, the absolute error is shown. For randomized methods, the root mean squared error from 300 independent replications is shown. The upper left panel shows, from top to bottom, the error for unscrambled, random linear scrambled and ASM scrambled Faure points. The Faure points lie very close to the $O(n^{-1})$ reference line, with no apparent evidence of a logarithmic factor. The matrix scrambled points are close to the $O(n^{-3/2})$ reference line. The ASM scrambled points seem to follow $O(n^{-3/2})$ at first, then approach the $O(n^{-2})$ reference before leveling out.

The upper right panel shows the same three methods, with a reflection incorporated. The curve for ASM scrambling keeps crossing the $n^{-2}$ reference line. The curve for random linear scrambling lies just below the $n^{-3/2}$ reference. The curve for reflection without scrambling has a prominent flat spot for $n \le 32{,}768$. Then it gets much better at 65,536.

The lower left panel shows the three methods with box symmetry. Here the curve for random linear scrambling lies between the references for $n^{-3/2}$ and $n^{-2}$ and ends up roughly parallel to the latter. The curve for ASM scrambling ends up below the $n^{-2}$ reference line. The curve for the box symmetrized Faure sequence follows the one for random linear scrambling, but has an error that is not monotone in $n$.

For each kind of symmetry, the ASM scrambling seems to give the best results on this function. The lower right panel shows all three ASM methods.

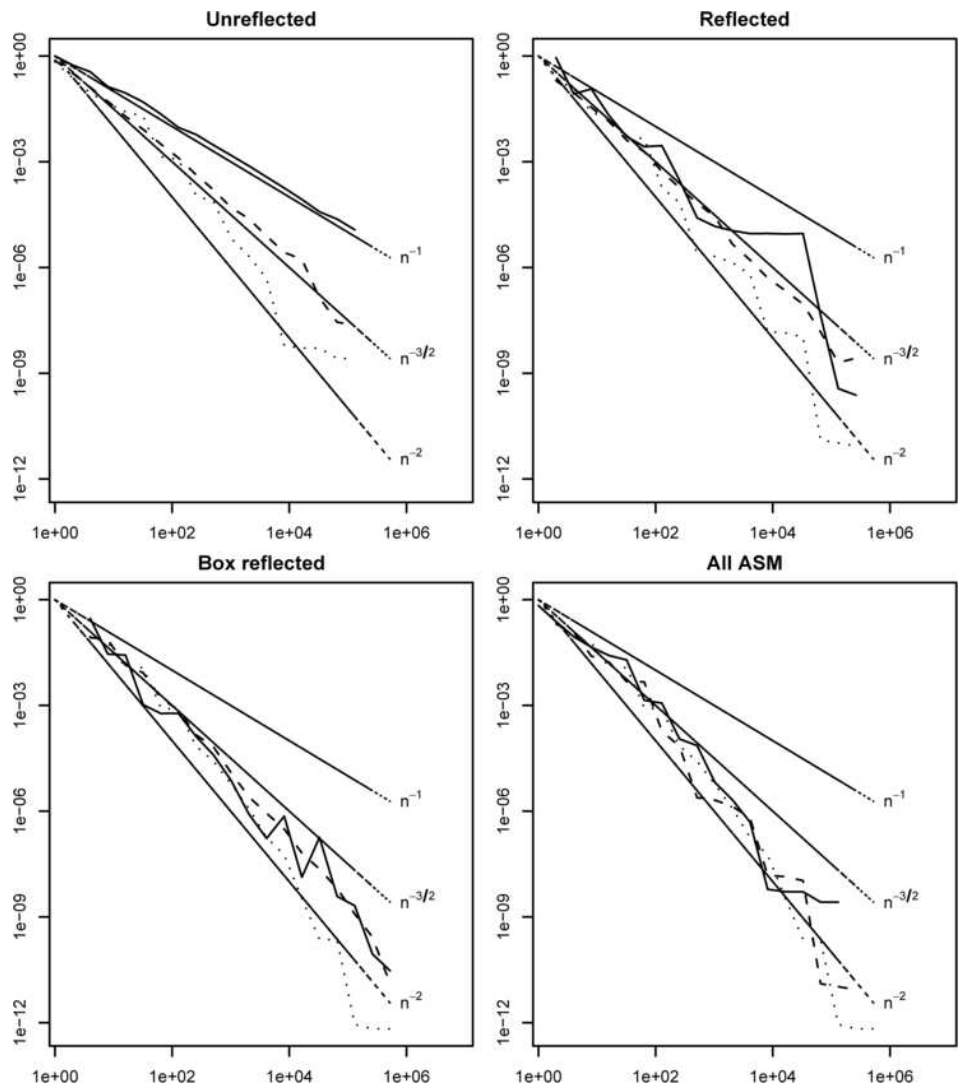

Fig. 2. *Shown are absolute errors for the Faure sequence and sample RMSEs from 300 replications for scrambled versions, in the quadrature example of Section 4. The lower right panel is for ASM scrambling: unreflected (solid), reflected (dashed) and box (dotted). The other panels depict unscrambled (solid), linearly scrambled (dashed) and ASM scrambled (dotted) results. All panels have reference lines proportional to labeled powers of n.*

From top to bottom at the right of that panel they are for the original points, reflected points and boxed points.

From this example it is clear that reflection strategies have potential to bring improvements and may even yield a rate better than $O(n^{-3/2})$. There are also some prominent flat spots and reversals in the errors. In the next sections we investigate box reflections and show that it can improve the error rate.

**5. Variance for scrambled digital nets.** The error rate analysis for box reflection of scrambled digital nets builds on the analysis for unreflected scrambled nets. This section recaps some needed material for completeness, widens the generality, and corrects an error in the original proof.

We begin by recapping a base $b$ Haar wavelet multiresolution of functions on $[0,1)^d$. For more details, see [20] and [21].

First define the univariate mother wavelets for $x \in \mathbb{R}$:
$$\psi_c(x) = b^{1/2} 1_{\lfloor bx \rfloor = c} - b^{-1/2} 1_{\lfloor x \rfloor = 0}, \qquad c = 0, 1, \ldots, b-1.$$

The familiar ($b = 2$) Haar wavelet decomposition only needs one mother wavelet because it has $\psi_0 = -\psi_1$. The general setting considered here requires more than one mother wavelet. Next, for nonnegative integers $k$ and $t < b^k$ define dilated and translated versions for $x \in [0,1)$,
$$\begin{aligned}\psi_{ktc}(x) &= b^{k/2} \psi_c(b^k x - t), \\ &= b^{(k+1)/2} 1_{\lfloor b^{k+1} x \rfloor = bt + c} - b^{(k-1)/2} 1_{\lfloor b^k x \rfloor = t} \\ &\equiv b^{(k+1)/2} N_{k,t,c}(x) - b^{(k-1)/2} W_{k,t}(x).\end{aligned}$$

The functions $N$ and $W$ are indicators of relatively narrow and wide intervals respectively, where the base $b$ is understood. Each $\psi_{ktc}$ is a narrow rectangular spike minus another one that is $b$ times as wide, but $1/b$ times as high.

The wavelets for $d \geq 1$ are tensor products of functions of the form $\psi_{ktc}$. For $u \subseteq 1 : d$, let $\kappa$ be a $|u|$-vector of integers $k_j \geq 0$ for $j \in u$. Similarly, let $\tau$ be a $|u|$-vector of nonnegative integers $t_j < b^{k_j}$ for $j \in u$. Notice that for $\kappa$ to be well defined a set $u$ must be understood, and $\tau$ depends similarly on both $u$ and $\kappa$. To avoid cluttered notation, we do not write $\kappa(u)$ or $\tau(u, \kappa)$. The $d$ variate Haar wavelets in base $b$ take the form
$$\psi_{u\kappa\tau\gamma}(x) = \prod_{j \in u} \psi_{k_j t_j c_j}(x^j),$$
with $\psi_{\{\}()()()}(x) = 1$ by convention.

The multiresolution of $f \in L^2[0,1)^d$ is
$$f(x) = \sum_u \sum_\kappa \sum_\tau \sum_\gamma \langle \psi_{uktg}, f \rangle \psi_{uktg}(x),$$
$$\langle \psi_{uktg}, f \rangle = \int \psi_{uktg}(x) f(x) \, dx,$$
where each summation is over all possible values for its argument, beginning with all subsets $u$ of $\{1, \ldots, d\}$.

It is convenient to write $f(x) = \sum_u \sum_\kappa \nu_{u\kappa}(x)$, where
$$\nu_{u\kappa}(x) = \sum_\tau \sum_\gamma \langle \psi_{u\kappa\tau\gamma}, f \rangle \psi_{u\kappa\tau\gamma}(x).$$

The function $\nu_{uk}(x)$ is a step function constant within elementary intervals of the form $\mathbb{B}_{u,\kappa,\tau}$.

If $x_1, \ldots, x_n$ are obtained by making a nested uniform (or random linear or I-binomial) scramble of points $a_1, \ldots, a_n \in [0,1)^d$ in base $b$, then the variance of $\hat{I} = n^{-1} \sum_{i=1}^n f(x_i)$ is

(13)
$$\frac{1}{n} \sum_{|u|>0} \sum_\kappa \Gamma_{u,\kappa} \sigma_{u,\kappa}^2,$$

where
$$\sigma_{u,\kappa}^2 = \int \nu_{u,\kappa}(x)^2 \, dx,$$

and the "gain coefficients" are given by
$$\Gamma_{u,\kappa} = \frac{1}{n(b-1)^{|u|}} \sum_{i=1}^n \sum_{i'=1}^n \prod_{j \in u} (b 1_{\lfloor b^{k_j+1} a_i^j \rfloor = \lfloor b^{k_j+1} a_{i'}^j \rfloor} - 1_{\lfloor b^{k_j} a_i^j \rfloor = \lfloor b^{k_j} a_{i'}^j \rfloor}).$$

From the "multiresolution ANOVA," $\sigma^2 = \sum_u \sum_\kappa \sigma_{u,\kappa}^2$. Therefore, the variance of ordinary Monte Carlo sampling has the form (13) with all $\Gamma_{u,\kappa} = 1$. The variance reduction from randomized nets arises from $\Gamma_{u,\kappa} \ll 1$ for some $u$ and $\kappa$ without allowing $\Gamma_{u,\kappa} \gg 1$ for any $u$ and $\kappa$. In particular, if $a_1, \ldots, a_n$ are a $(\lambda, q, m, d)$-net in base $b$, then $\Gamma_{u,\kappa} = 0$ if $m - q \geq |u| + |\kappa|$.

THEOREM 2. *Let $a_1, \ldots, a_n$ be a $(0, m, d)$-net in base $b \geq 2$. Then*
$$0 \leq \Gamma_{u,\kappa} \leq \left(\frac{b}{b-1}\right)^{\min(d-1,m)} \leq \left(\frac{b}{b-1}\right)^{b-1} \leq e \doteq 2.718.$$

*Let $a_1, \ldots, a_n$ be a $(\lambda, 0, m, d)$-net in base $b \geq 2$. Then*
$$0 \leq \Gamma_{u,\kappa} \leq e + 1 \doteq 3.718.$$

*Let $a_1, \ldots, a_n$ be a $(\lambda, q, m, d)$-net in base $b \geq 2$. Then*
$$0 \leq \Gamma_{u,\kappa} \leq b^q \left(\frac{b}{b-1}\right)^{d-1}.$$

PROOF. The first part is from [20], the second is from [21], and the third is from [22]. □

Theorem 2 shows some upper bounds on gain coefficients for nets. Sharper, but more complicated bounds are available from intermediate stages of the proofs, particularly the ones in [22]. Still sharper bounds are available in [18] and in [31].

5.1. *Scrambled net variance for smooth functions.* There is an error in the way that the $O(\cdots)$ terms are gathered in Lemma 1 of [21]. This section repairs the proof of the $O(n^{-3}\log(n)^{d-1})$ result for the variance of scrambled net integrals of smooth functions. In the process, a more general result is obtained, using a weaker definition of smoothness than in the original paper, and covering nets with nonzero quality parameter and relaxed versions of $(\lambda, q, m, d)$-nets.

The proof follows the lines of [21]. Lemmas 2 and 3 here replace Lemmas 1 and 2 there, respectively.

LEMMA 2. *Suppose that $f$ is a smooth function on $[0,1]^d$. For $b \geq 2$ and $u \subseteq \{1, \ldots, d\}$, let $\kappa$ and $\tau$ be $|u|$-tuples of nonnegative integers with components $k_j$ and $t_j < b^{k_j}$ for $j \in u$. Then*

$$(14) \qquad |\langle f, \psi_{u\kappa\tau\gamma}\rangle| \leq \left(\frac{b-1}{b}\right)^{|u|} b^{-(3|\kappa|+|u|)/2} \sup_{x \in \mathbb{B}_{u,\kappa,\tau}} |\partial^u f_u(x)|.$$

PROOF. From the definitions,

$$\langle f, \psi_{u\kappa\tau\gamma}\rangle$$
$$= \langle f_u, \psi_{u\kappa\tau\gamma}\rangle$$
$$= b^{-(|\kappa|+|u|)/2} \int f_u(x)\psi_{u\kappa\tau\gamma}(x)\,dx$$
$$(15) \qquad = b^{-(|\kappa|+|u|)/2} \int f_u(x) \prod_{j\in u} b^{k_j+1}(N_{k_jt_jc_j}(x^j) - b^{-1}W_{k_jt_j}(x^j))\,dx.$$

Next, $f_u(x)$ depends on $x$ only through $x^u$. Applying (7) to $f_u$, we may write

$$(16) \qquad f_u(x) = \sum_{v \subseteq u} \int_{[\mathsf{c}_{u\kappa\tau}^v, x^v]} \partial^v f_u(\mathsf{c}_{u\kappa\tau}^{-v} : y^v)\,dy^v.$$

If $v \neq u$, then the corresponding term in (16) does not depend on $x_{u-v}$ and is therefore orthogonal to $N_{k_jt_jc_j}(x^j) - b^{-1}W_{k_jt_j}(x^j)$ for $j \in u - v$. Accordingly, we may replace $f_u$ in (15) by the $v = u$ term from (16). Also, the integrand in (15) vanishes for $x \notin \mathbb{B}_{u\kappa\tau}$. Putting these together, we find that $b^{(|\kappa|+|u|)/2}\langle f, \psi_{u\kappa\tau\gamma}\rangle$ equals

$$\int \int_{[\mathsf{c}_{u\kappa\tau}^u, x^u]} \partial^u f_u(\mathsf{c}_{u\kappa\tau}^{-u} : y^u)\,dy^u \prod_{j\in u} b^{k_j+1}(N_{k_jt_jc_j}(x^j) - b^{-1}W_{k_jt_j}(x^j))\,dx$$

$$\leq \sup_{x^u \in \mathbb{B}_{u\kappa\tau}} \left|\int_{[\mathsf{c}_{u\kappa\tau}^u, x^u]} \partial^u f_u(\mathsf{c}_{u\kappa\tau}^{-u} : y^u)\,dy^u\right|$$

$$\times \int \prod_{j \in u} b^{k_j+1} |N_{k_j t_j c_j}(x^j) - b^{-1} W_{k_j t_j}(x^j)| \, dx$$

$$= (2 - 2/b)^{|u|} \sup_{x^u \in \mathbb{B}_{u\kappa\tau}} \left| \int_{[\mathsf{c}^u_{u\kappa\tau}, x^u]} \partial^u f_u(\mathsf{c}^{-u}_{u\kappa\tau} : y^u) \, dy^u \right|.$$

By Lemma 1, $\partial^u f_u$ is continuous, and so by the mean value theorem, there is a point $z \in \mathbb{B}_{u\kappa\tau}$ with

$$\left| \int_{[\mathsf{c}^u_{u\kappa\tau}, x^u]} \partial^u f_u(\mathsf{c}^{-u}_{u\kappa\tau} : y^u) \, dy^u \right| = \mathrm{Vol}(\mathrm{rect}[\mathsf{c}^u_{u\kappa\tau}, x^u]) |\partial^u f_u(z)|$$

$$\leq 2^{-|u|} b^{-|\kappa|} |\partial^u f_u(z)|.$$

The factor $b^{-|\kappa|}$ is the volume of a $|u|$-dimensional elementary interval containing both $\mathsf{c}^u_{u\kappa\tau}$ and $x^u$. The factor $2^{-|u|}$ arises because $\mathsf{c}^u_{u\kappa\tau}$ is at the center of this elementary interval and $x^u$ is in some sub-interval defined by $\mathsf{c}^u_{u\kappa\tau}$ and one of the corners of that elementary interval. Finally,

$$|\langle f, \psi_{u\kappa\tau\gamma} \rangle| \leq (1 - 1/b)^{|u|} b^{-(3|\kappa|+|u|)/2} \sup_{z \in \mathbb{B}_{u\kappa\tau}} |\partial^u f_u(z)|. \qquad \square$$

LEMMA 3. *Under the conditions of Lemma 2,*

$$(17) \qquad \sigma^2_{u\kappa} \leq 2^{|u|} \left(\frac{b-1}{b}\right)^{3|u|} b^{-2|\kappa|} \|\partial^u f_u\|^2_\infty.$$

PROOF. The supports of $\psi_{u\kappa\tau\gamma}$ and $\psi_{u\kappa\tau'\gamma'}$ are disjoint unless $\tau = \tau'$, and so

$$\nu^2_{u\kappa}(x) = \sum_\tau \sum_\gamma \sum_{\gamma'} \langle f, \psi_{u\kappa\tau\gamma} \rangle \langle f, \psi_{u\kappa\tau\gamma'} \rangle \psi_{u\kappa\tau\gamma}(x) \psi_{u\kappa\tau\gamma'}(x).$$

Now

$$\sigma^2_{u\kappa} = \int \nu^2_{u\kappa}(x) \, dx$$

$$= \sum_\tau \sum_\gamma \sum_{\gamma'} \langle f, \psi_{u\kappa\tau\gamma} \rangle \langle f, \psi_{u\kappa\tau\gamma'} \rangle \int \psi_{u\kappa\tau\gamma}(x) \psi_{u\kappa\tau\gamma'}(x) \, dx$$

$$= \sum_\tau \sum_\gamma \sum_{\gamma'} \langle f, \psi_{u\kappa\tau\gamma} \rangle \langle f, \psi_{u\kappa\tau\gamma'} \rangle \prod_{j \in u} (1_{c_j = c'_j} - b^{-1})$$

$$\leq \left(\frac{b-1}{b}\right)^{2|u|} b^{-3|\kappa|-|u|} \sum_\tau \sup_{z \in \mathbb{B}_{u\kappa\tau}} |\partial^u f_u(z)|^2 \sum_\gamma \sum_{\gamma'} \prod_{j \in u} |1_{c_j = c'_j} - b^{-1}|$$

$$\leq \left(\frac{b-1}{b}\right)^{2|u|} b^{-3|\kappa|-|u|} \left(\sum_\tau \sup_{z \in \mathbb{B}_{u\kappa\tau}} |\partial^u f_u(z)|^2\right) \left(\sum_{c=0}^{b-1} \sum_{c'=0}^{b-1} |1_{c_j = c'_j} - b^{-1}|\right)^{|u|}$$

$$= 2^{|u|} \left(\frac{b-1}{b}\right)^{3|u|} b^{-3|\kappa|} \sum_\tau \sup_{z \in \mathbb{B}_{u\kappa\tau}} |\partial^u f_u(z)|^2$$

$$\leq 2^{|u|} \left(\frac{b-1}{b}\right)^{3|u|} b^{-2|\kappa|} \|\partial^u f_u\|_\infty^2. \qquad \square$$

THEOREM 3. *Let $x_1$ through $x_n$ be the points of a randomized relaxed $(\lambda, q, m, d)$-net in base $b$. Suppose that as $n \to \infty$ with $\lambda$ and $q$ fixed, that all of the gain coefficients of the net satisfy $\Gamma_{u\kappa} \leq G < \infty$. Then for smooth $f$,*

$$V(\hat{I}) = O\left(\frac{(\log n)^{d-1}}{n^3}\right).$$

PROOF. If $|\kappa| + |u| \leq m - q$, then the digital net property of $x_1, \ldots, x_n$ yields $\Gamma_{u\kappa} = 0$. Otherwise, we have $\Gamma_{u\kappa} \leq G$, and so

$$V(\hat{I}) \leq \frac{G}{n} \sum_{|u|>0} \sum_{|\kappa|>(m-q-|u|)_+} \sigma_{u\kappa}^2$$

$$\leq \frac{G}{n} \sum_{|u|>0} \sum_{|\kappa|>(m-q-|u|)_+} 2^{|u|} \left(\frac{b-1}{b}\right)^{3|u|} \|\partial^u f_u\|_\infty^2 b^{-2|\kappa|}$$

(18)
$$\leq \frac{G'}{n} \sum_{|u|>0} \sum_{|\kappa|>(m-q-|u|)_+} b^{-2|\kappa|},$$

where

$$G' = G 2^{|u|} \left(\frac{b-1}{b}\right)^{3|u|} \max_{|u|>0} \|\partial^u f_u\|_\infty^2.$$

Because we are interested in the limit as $m \to \infty$, we may suppose that $m > d + q$. For such large $m$,

$$\sum_{|\kappa|>(m-q-|u|)_+} b^{-2|\kappa|} = \sum_{r=m-q-|u|+1}^\infty b^{-2r} \binom{r+|u|-1}{|u|-1},$$

where the binomial coefficient is the number of $|u|$-vectors $\kappa$ of nonnegative integers that sum to $r$. Making the substitution $s = r - m + q + |u|$,

$$\sum_{|\kappa|>(m-q-|u|)_+} b^{-2|\kappa|}$$

$$= b^{-2m+2q+2|u|} \sum_{s=1}^\infty b^{-2s} \binom{s+m-q-1}{|u|-1}$$

$$\leq \frac{\lambda^2}{n^2} \frac{b^{2q+2|u|}}{(|u|-1)!} \sum_{s=1}^\infty b^{-2s} (s+m-q-1)^{|u|-1}$$

$$\leq \frac{\lambda^2}{n^2} \frac{b^{2q+2|u|}}{(|u|-1)!} \sum_{s=1}^{\infty} b^{-2s} \sum_{j=0}^{|u|-1} \binom{|u|-1}{j} s^j (m-q-1)^{|u|-1-j}$$

$$\leq \frac{\lambda^2}{n^2} b^{2(q+|u|-1)} \sum_{j=0}^{|u|-1} \frac{(m-q-1)^{|u|-1-j}}{j!(|u|-1-j)!} \sum_{s=1}^{\infty} b^{-2(s-1)} s^j$$

$$\leq \frac{\lambda^2}{n^2} |u| b^{2(q+|u|-1)} m^{|u|-1} \sum_{s=1}^{\infty} b^{-2(s-1)} s^{|u|-1}$$

(19) $$= O(n^{-2} \log(n)^{|u|-1}),$$

because the infinite sum converges, $m \leq \log_b(n)$ and $|u| \leq d$. The theorem follows upon substituting the bound (19) into (18). $\square$

**6. Scrambled net variance with box folding.** This section investigates the effects of reflection schemes on scrambled net variance. Reflections are written as $\mathcal{R}_\rho$, where $\rho$ is a $d$ vector of integers $r_j \geq -1$. As before, we let $\kappa$ denote a scale for the multiresolution analysis.

In Section 5.1 the coefficients $\langle f, \psi_{u\kappa\tau\gamma} \rangle$ are bounded in terms of mixed partial derivatives of $f$ taken once with respect to each component $x^j$ for $j \in u$. Reflection is a piece-wise differentiable operation. The function $\mathcal{R}_\rho(x)$ is discontinuous at $x$ if $x^j = tb^{-r_j}$ holds for some $j$ with $r_j > 0$ and some positive integer $t < b^{r_j}$. In the interior of the pieces, reflection of $x^j$ reverses the sign of the derivative with respect to $x^j$. This sign reversal can be exploited to produce a cancellation effect that reduces a bound on $\langle f, \psi_{u\kappa\tau\gamma} \rangle$.

To simplify some expressions, we define the composite function $f^\rho$ by $f^\rho(x) = f(\mathcal{R}_\rho(x))$. At almost all points $x \in [0,1]^d$ the chain rule gives

(20) $$\partial^u f^\rho(x) = (-1)^{\operatorname{sgn}(\rho)} \partial^u f(\mathcal{R}_\rho(x)),$$

where $\operatorname{sgn}(\rho) = \sum_{j=1}^d 1_{r_j \geq 0}$ counts the number of reflections in $\rho$. The factor $\partial^u f(\mathcal{R}_\rho(x))$ in the right-hand side of (20) is the partial derivative of $f$, evaluated at the point $z = \mathcal{R}_\rho(x)$, and not the partial derivative of $f \circ \mathcal{R}_\rho$ evaluated at $x$, which appears on the left-hand side.

DEFINITION 11. In $d$ dimensions, a box folding scheme is an average of $2^d$ reflections as described below. Start with $\rho = (r_1, \ldots, r_d)$, where each $r_j \geq 0$. For $\ell = 0, \ldots, 2^d - 1$, let $\rho_\ell$ be the $d$ vector of integer components $r_{\ell j} \in \{r_j, -1\}$ with $r_{\ell j} = r_j$ if and only if the $j$th base 2 digit of $\ell$ is one. Then the box fold scheme is

$$\widetilde{I} = \frac{1}{2^d} \sum_{\ell=0}^{2^d-1} \frac{1}{n} \sum_{i=1}^n f^{\rho_\ell}(x_i) = \frac{1}{n} \sum_{i=1}^n \widetilde{f}(x_i),$$

where $\widetilde{f}(x) = 2^{-d} \sum_{\ell=0}^{2^d-1} f^{\rho_\ell}(x)$.

Sometimes it is more convenient to index the reflections by $2^d$ subsets $v \subseteq \{1, \ldots, d\}$. Let $v = v(\ell)$ denote the subset where $j \in v$ if and only if the $j$th binary digit of $\ell$ is a one. Taking $\rho_v$ to mean $\rho_\ell$ where $v = v(\ell)$, we may write $\widetilde{f}(x) = 2^{-d} \sum_{v \subseteq 1:d} f^{\rho_v}(x)$. From the definition of $v$, we find that $\mathrm{sgn}(\rho_v) = (-1)^{|v|}$.

To get ANOVA components of $\widetilde{f}$, we need the ANOVA components of $f^\rho$. Lemma 4 below shows that reflection commutes with the operation of taking ANOVA components.

LEMMA 4. *Let $f$ be an $L^2$ function on $[0,1]^d$. Let $f^\rho(x) = f(\mathcal{R}_\rho(x))$, where $\rho$ is a $d$ vector of integers $r_j \geq -1$ for $j = 1, \ldots, d$. Let $u \subseteq \{1, \ldots, d\}$. Then*

$$f^\rho_u(x) = f_u(\mathcal{R}_\rho(x)). \tag{21}$$

PROOF. The proof follows by induction on $|u|$. □

The bounds for $\langle f, \psi_{u\kappa\tau\gamma} \rangle$ in Section 5.1 made use of differentiability of $f$, which we cannot assume for $f^\rho$. The derivation as far as equation (15) does follow for $f^\rho$ and so $\langle f^\rho, \psi_{u\kappa\tau\gamma} \rangle$ equals

$$b^{-(|\kappa|+|u|)/2} \int f^\rho_u(x) \prod_{j \in u} b^{k_j+1}(N_{k_j t_j c_j}(x^j) - b^{-1}W_{k_j t_j}(x^j))\, dx. \tag{22}$$

The next step in the derivation of bounds for $\langle f, \psi_{u\kappa\tau\gamma} \rangle$ required $\partial^u f$ at points of $\mathbb{B}_{u\kappa\tau}$, and $\partial^u f^\rho$ does not necessarily exist.

The setting is simplest if the scale $\kappa$ is finer than the reflection $\rho$. Suppose that $u = \{1, \ldots, d\}$ and that $k_j \geq r_j$ for $j = 1, \ldots, d$. This specifically includes cases with $r_j = -1$ that designate no reflection for component $j$. Then, for smooth $f$, $\partial^u f_\rho$ is uniformly continuous on the interior of $\mathbb{B}_{u\kappa\tau}$. Letting $\mathsf{c}_{u\kappa\tau}$ be the center of $\mathbb{B}_{u\kappa\tau}$ as before, we find that

$$\langle f^\rho, \psi_{u\kappa\tau\gamma} \rangle$$
$$= \int \int_{[\mathsf{c}^u_{u\kappa\tau}, x^u]} \partial^u f^\rho_u(\mathsf{c}^{-u}_{u\kappa\tau} : y^u)\, dy^u\, \psi_{u\kappa\tau\gamma}(x)\, dx$$
$$= (-1)^{\mathrm{sgn}(\rho)} \int \int_{[\mathsf{c}^u_{u\kappa\tau}, x^u]} \partial^u f_u(\mathsf{c}^{-u}_{u\kappa\tau} : \mathcal{R}_\rho(y)^u)\, dy^u\, \psi_{u\kappa\tau\gamma}(x)\, dx$$
$$= (-1)^{\mathrm{sgn}(\rho)} \int \int_{[\mathsf{c}_{u\kappa\tau}, x]} \partial^u f_u(\mathcal{R}_\rho(y))\, dy\, \psi_{u\kappa\tau\gamma}(x)\, dx, \tag{23}$$

where at the last step we use $u = 1\!:\!d$ and $-u = \varnothing$.

LEMMA 5. *Suppose that $f$ is a doubly smooth function on $[0,1]^d$. Let $\rho = (r_1, \ldots, r_d)$ with integers $r_j \geq 0$. Take $|\rho| = \sum_{j=1}^d r_j$, and let $\widetilde{f}$ be defined by the box folding scheme of Definition 11. For $b \geq 2$ and $u = \{1, \ldots, d\}$, let $\kappa$, $\tau$ and $\gamma$ be d-tuples of nonnegative integers with components $k_j \geq r_j$, $t_j < b^{k_j}$, and $c_j < b$ respectively, for $j = 1, \ldots, d$. Then*

$$(24) \qquad |\langle \widetilde{f}, \psi_{u\kappa\tau\gamma} \rangle| \leq b^{-|\rho|} \left( \frac{b-1}{b} \right)^{-d} b^{-(3|\kappa|+|u|)/2} \|\partial^{u,u} f_u\|_\infty.$$

PROOF. Because $\kappa$ is on a finer scale than all of the reflections $\rho_\ell$, equation (23) holds for each of them. Therefore,

$$\langle \widetilde{f}, \psi_{u\kappa\tau\gamma} \rangle = \frac{1}{2^d} \int \int_{[\mathsf{c}_{u\kappa\tau}, x]} \sum_{v \subseteq 1:d} (-1)^{|v|} \partial^u f_u(\mathcal{R}_{\rho_v}(y)) \, dy \, \psi_{u\kappa\tau\gamma}(x) \, dx.$$

For $y \in [0,1]^d$, let $\mathsf{k} = \mathsf{k}(y) \in [0,1]^d$ be the center point through which the reflection $\mathcal{R}_\rho$ with $\rho = (r_1, \ldots, r_d)$ operates on $y$. That is, $\mathsf{k}^j = b^{-r_j}(\lfloor b^{r_j} y^j \rfloor + 1/2)$. Because $\kappa$ is finer than $\rho$, the same center $\mathsf{k}$ applies for all $y \in [\mathsf{c}_{u\kappa\tau}, x]$. Then the $j$th component of $\mathcal{R}_{\rho_v}(y)$ is $2\mathsf{k}^j - y^j$ if $j \in v$ and is $y^j$ otherwise. Therefore,

$$\sum_{v \subseteq u} (-1)^{|v|} \partial^u f_u((2\mathsf{k}-y)^v : y^{-v}) = \mathrm{Vol}(\mathrm{rect}[y, 2\mathsf{k}-y]) \partial^{u,u} f_u(z),$$

where $z = z(y) \in \mathrm{rect}[y, 2\mathsf{k}-y]$. The volume of $\mathrm{rect}[y, 2\mathsf{k}-y]$ is at most $b^{-|\rho|}$ and so following the argument from Lemma 2,

$$\langle \widetilde{f}, \psi_{u\kappa\tau\gamma} \rangle \leq (1 - 1/b)^{-d} b^{-|\rho|} b^{-(3|\kappa|+|u|)/2} \|\partial^{u,u} f_u\|_\infty. \qquad \square$$

The factor $b^{-|\rho|}$ in (24) underlies the improvement that a box reflection can bring. For a scrambled $(\lambda, q, m, d)$-net in base $b$, if we choose $\rho$ so that $|\rho| = m - q$, then the coefficients $\langle \widetilde{f}, \psi_{u\kappa\tau\gamma} \rangle$ with $\kappa$ finer than $\rho$ are $O(b^{-3|\kappa|/2-|\rho|})$ instead of $O(b^{-3|\kappa|/2})$. Coarse terms with $|\kappa| + |u| \leq m - q$ do not contribute to the error, so the dominant error terms have $|\kappa| + |u| = m - q + 1$. In the next theorem we will deal with those terms by taking $|\rho| = m - q$. Choosing $|\rho| = m - q$, the largest contributing coefficients are $O(b^{-3|\kappa|/2-|\rho|}) = O(b^{-3m/2-m}) = O(n^{-5/2})$ instead of $O(b^{-3|\kappa|/2}) = O(b^{-3m/2}) = O(n^{-3/2})$. Following the derivation in Section 5.1, the terms $\sigma_{u\kappa}^2$ are then of order $O(b^{-3m}) = O(n^{-3})$ instead of $O(b^{-2m}) = O(n^{-2})$ and so each of them contributes $O(n^{-4})$ to the variance instead of $O(n^{-3})$. The variance under box folding does not generally end up as $O(n^{-4+\varepsilon})$ though, because there are also contributions from terms $\kappa$ where $\kappa$ is not finer than $\rho$.

THEOREM 4. *Let $x_1$ through $x_n$ be points of a randomized relaxed $(\lambda, q, m, d)$-net in base $b$. Suppose that the quality parameter $q$ remains fixed as $n$ tends to infinity through values $\lambda b^m$ for fixed $\lambda$ and that none of the gain coefficients of the net is larger than $G < \infty$. Then for doubly smooth $f$, under box folding by $\rho = (r_1, \ldots, r_d)$ where*

$$r_j = \begin{cases} \lfloor (m-q)/d \rfloor + 1, & j \leq (m-q) - d\lfloor (m-q)/d \rfloor \\ \lfloor (m-q)/d \rfloor, & \text{otherwise,} \end{cases}$$

*we find that*

$$V(\widetilde{I}) = O\left(\frac{(\log n)^{d-1}}{n^{3+2/d}}\right)$$

*as $n \to \infty$.*

PROOF. First we consider coefficients $\langle \widetilde{f}, \psi_{u\kappa\tau\gamma} \rangle$ for the highest order subset $u = \{1, \ldots, d\}$. Let $w = w(\kappa) = \{j \in u \mid k_j \geq r_j\}$. If $w = \varnothing$, then $\sum_{j \in u} k_j \leq \sum_{j \in u}(r_j - 1) = m - q - d$. Then $|\kappa| + |u| = m - q$, so that $\Gamma_{u\kappa} = 0$ by the balance property of the digital net. Therefore, we restrict attention to $w$ with $|w| > 0$. Lemma 5 treated the case with $w = u$ and with $\kappa$ finer than $\rho$.

For $x$ in the support of $\psi_{u\kappa\tau\gamma}$, the function $\widetilde{f}$ is differentiable with respect to $x^j$ for $j \in w$. We may apply equation (7) to each $f^{\rho_v}$, keeping only the $\partial^w$ term because the others are orthogonal to $\psi_{u\kappa\tau\gamma}$. The result shows that $2^d \langle \widetilde{f}, \psi_{u\kappa\tau\gamma} \rangle$ is

$$\int \sum_{v \subseteq 1:d} f_u^{\rho_v}(x) \psi_{u\kappa\tau\gamma}(x) \, dx$$

$$= \int \sum_{v \subseteq 1:d} \int_{[\mathsf{c}_{w\kappa\tau}^w, x^w]} \partial^w f_u^{\rho_v}(x^{-w} : y^w) \, dy^w \, \psi_{u\kappa\tau\gamma}(x) \, dx$$

(25) $$= \int \sum_{v_1 \subseteq -w} \int_{[\mathsf{c}_{w\kappa\tau}^w, x^w]} \sum_{v_2 \subseteq w} \partial^w f_u^{\rho_{v_1 \cup v_2}}(x^{-w} : y^w) \, dy^w \, \psi_{u\kappa\tau\gamma}(x) \, dx,$$

after decomposing $v$ into its intersections $v_1$ and $v_2$ with $w$ and $-w$ respectively.

The summation inside of (25) may be written as

$$\sum_{v_2 \subseteq w} (-1)^{|v_2|} \partial^w f_u(\mathcal{R}_{\rho_{v_1 \cup v_2}}(x)^{-w} : \mathcal{R}_{\rho_{v_1 \cup v_2}}(y)^w)$$

$$= \operatorname{Vol}(\operatorname{rect}[y^w, 2\mathsf{k}_{v_1}^w - y^w]) \partial^{w,w} f_u(\mathcal{R}_{\rho_{v_1 \cup v_2}}(x)^{-w} : z^w),$$

where for $j \in w$, $\mathsf{k}_{v_1}^j = b^{-r_j}(\lfloor b^{r_j} y^j \rfloor + 1/2)$ and $z^w \in \operatorname{rect}[y^w, 2\mathsf{k}_{v_1}^w - y^w]$. Because $\operatorname{Vol}(\operatorname{rect}[y^w, 2\mathsf{k}_{v_1}^w - y^w]) \leq b^{-\sum_{j \in w} r_j}$, we find that box reflection results in a coefficient $\langle \widetilde{f}, \psi_{u\kappa\tau\gamma} \rangle$ with an upper bound on the order of $b^{-\sum_{j \in w} r_j}$ smaller than the bound for $\langle f, \psi_{u\kappa\tau\gamma} \rangle$.

This coefficient reduction is $b^{-\sum_{j \in w} r_j} = O(b^{-m|w|/d}) = O(n^{-|w|/d})$. Because we only need to consider nonempty $w$, the reduction is $O(n^{-1/d})$. The effect is to reduce the bound for $\sigma^2_{u\kappa}$ by $O(n^{-2/d})$ and then the same counting argument as in Theorem 3 shows that the contribution of $f_u$ to the variance is $O((\log n)^{d-1}/n^{3+2/d})$.

Now consider variance contribution of $f_v$ for $v \subset \{1, \ldots, d\}$ with $1 \le |v| < d$. The sum $(1/n) \sum_{i=1}^n \widetilde{f}_v(x_i)$ is a box fold of a scrambled relaxed $(\lambda b^{d-|v|}, q, m, |v|)$-net in base $b$ for estimating the mean of the fully $|v|$-dimensional function $g(x^v) = f_v(x^v : 0^{-v})$ obtained by ignoring the $-v$ components of $x$. Accordingly, it makes a variance contribution that is $O((\log n)^{|v|-1}/n^{3+2/|v|})$. The variance of the sum cannot be of higher order than $O((\log n)^{d-1}/n^{3+2/d})$. □

**7. Discussion.** In this paper we have seen that scrambled net quadrature can be profitably combined with antithetic sampling to reduce variance. This result then fits in with the work of [12] who combined quasi-Monte Carlo with control variates and [27] and [3] who both looked at quasi-Monte Carlo in combination with importance sampling. The best numerical results were for ASM scrambling combined with box reflections, but we have no theoretical results for that combination.

The foldings of scrambled nets studied here may also be viewed as a hybrid of digital nets and a monomial cubature rule. The $2^d$-fold symmetry used by box folding takes each sample point in the net and uses it to generate the points of a cubature. It is one of many cubature rules that might be made to work with digital nets. For background and catalogues of cubature rules, see [4, 5] and [28].

The conclusions of Theorems 3 and 4 both hold if $\lambda$ and $q$ are allowed to fluctuate as $n$ increases, so long as both remain below finite upper bounds.

A larger improvement from local antithetic sampling may be possible if we can identify $s < d$ input variables that are much more important than the others, and apply reflections only to them. In some cases we can even re-engineer the integrand to make a small number of variables much more important than they are in the nominal encoding. For an example of such a technique with an integrand with respect to a high dimensional geometric Brownian motion, see [1] and [2]. Many more examples are presented in [9].

**Acknowledgments.** I thank Harald Niederreiter, Christiane Lemieux, a referee and the Associate Editor for helpful comments.

## REFERENCES

[1] ACWORTH, P., BROADIE, M. and GLASSERMAN, P. (1997). A comparison of some Monte Carlo techniques for option pricing. In *Monte Carlo and Quasi-Monte Carlo Methods'96* (H. Niederreiter, P. Hellekalek, G. Larcher and P. Zinterhof, eds.) 1–18. Springer, Berlin.

[2] CAFLISCH, R., MOROKOFF, E. W. and OWEN, A. B. (1997). Valuation of mortgage backed securities using Brownian bridges to reduce effective dimension. *J. Computational Finance* **1** 27–46.

[3] CHELSON, P. (1976). Quasi-random techniques for Monte Carlo methods. Ph.D. thesis, The Claremont Graduate School.

[4] COOLS, R. (1999). Monomial cubature rules since Stroud: A compilation—part 2. *J. Comput. Appl. Math.* **112** 21–27. MR1728449

[5] COOLS, R. and RABINOWITZ, P. (1993). Monomial cubature rules since Stroud: A compilation. *J. Comput. Appl. Math.* **48** 309–326. MR1252544

[6] EFRON, B. and STEIN, C. (1981). The jackknife estimate of variance. *Ann. Statist.* **9** 586–596. MR0615434

[7] FAURE, H. (1982). Discrépance de suites associées à un système de numération (en dimension $s$). *Acta Arith.* **41** 337–351. MR0677547

[8] FISHMAN, G. S. (2006). *A First Course in Monte Carlo.* Duxbury, Belmont, CA.

[9] FOX, B. L. (1999). *Strategies for Quasi-Monte Carlo.* Kluwer Academic, Boston.

[10] GLASSERMAN, P. (2004). *Monte Carlo Methods in Financial Engineering.* Springer, New York. MR1999614

[11] HABER, S. (1970). Numerical evaluation of multiple integrals. *SIAM Rev.* **12** 481–526. MR0285119

[12] HICKERNELL, F. J., LEMIEUX, C. and OWEN, A. B. (2005). Control variates for quasi-Monte Carlo (with discussion). *Statist. Sci.* **20** 1–31. MR2182985

[13] HLAWKA, E. (1961). Funktionen von beschränkter Variation in der Theorie der Gleichverteilung. *Ann. Mat. Pura Appl.* **54** 325–333. MR0139597

[14] HOEFFDING, W. (1948). A class of statistics with asymptotically normal distribution. *Ann. Math. Statist.* **19** 293–325. MR0026294

[15] L'ECUYER, P. and LEMIEUX, C. (2002). A survey of randomized quasi-Monte Carlo methods. In *Modeling Uncertainty*: *An Examination of Stochastic Theory, Methods, and Applications* (M. Dror, P. L'Ecuyer and F. Szidarovszki, eds.) 419–474. Kluwer Academic, New York.

[16] MATOUŠEK, J. (1998). *Geometric Discrepancy*: *An Illustrated Guide.* Springer, Heidelberg. MR1697825

[17] NIEDERREITER, H. (1992). *Random Number Generation and Quasi-Monte Carlo Methods.* SIAM, Philadelphia. MR1172997

[18] NIEDERREITER, H. and PIRSIC, G. (2001). The microstructure of $(t, m, s)$-nets. *J. Complexity* **17** 683–696. MR1881664

[19] OWEN, A. B. (1995). Randomly permuted $(t, m, s)$-nets and $(t, s)$-sequences. In *Monte Carlo and Quasi-Monte Carlo Methods in Scientific Computing* (H. Niederreiter and P. J.-S. Shiue, eds.) 299–317. Springer, New York. MR1445791

[20] OWEN, A. B. (1997). Monte Carlo variance of scrambled equidistribution quadrature. *SIAM J. Numer. Anal.* **34** 1884–1910. MR1472202

[21] OWEN, A. B. (1997). Scrambled net variance for integrals of smooth functions. *Ann. Statist.* **25** 1541–1562. MR1463564

[22] OWEN, A. B. (1998). Scrambling Sobol' and Niederreiter–Xing points. *J. Complexity* **14** 466–489. MR1659008

[23] Owen, A. B. (2003). Variance with alternative scramblings of digital nets. *ACM Trans. Modeling and Computer Simulation* **13** 363–378.

[24] Schürer, R. and Schmid, W. C. (2006). MinT: A database for optimal net parameters. In *Monte Carlo and Quasi-Monte Carlo Methods 2004* (H. Niederreiter and D. Talay, eds.) 457–469. Springer, Berlin. MR2208725

[25] Sloan, I. H. and Joe, S. (1994). *Lattice Methods for Multiple Integration*. Oxford Science Publications. MR1442955

[26] Sobol', I. M. (1967). The use of Haar series in estimating the error in the computation of infinite-dimensional integrals. *Dokl. Akad. Nauk SSSR* **8** 810–813. MR0215527

[27] Spanier, J. and Maize, E. H. (1994). Quasi-random methods for estimating integrals using relatively small samples. *SIAM Rev.* **36** 18–44. MR1267048

[28] Stroud, A. H. (1971). *Approximate Calculation of Multiple Integrals*. Prentice-Hall, Englewood Cliffs, NJ. MR0327006

[29] Takemura, A. (1983). Tensor analysis of ANOVA decomposition. *J. Amer. Statist. Assoc.* **78** 894–900. MR0727575

[30] Tezuka, S. and Faure, H. (2003). I-binomial scrambling of digital nets and sequences. *J. Complexity* **19** 744–757. MR2040428

[31] Yue, R. X. and Hickernell, F. J. (2002). The discrepancy and gain coefficients of scrambled digital nets. *J. Complexity* **18** 135–151. MR1895080

Department of Statistics
Stanford University
Stanford, California 94305
USA
E-mail: art@stat.stanford.edu